\documentclass{article}

\usepackage{arxiv_compact_authors}

\usepackage[utf8]{inputenc} 
\usepackage[T1]{fontenc}    
\usepackage{hyperref}       
\usepackage{url}            
\usepackage{booktabs}       
\usepackage{amsfonts}       
\usepackage{nicefrac}       
\usepackage{microtype}      
\usepackage{lipsum}
\usepackage{graphicx}
\graphicspath{ {./images/} }
\usepackage[numbers,sort&compress]{natbib}

\title{Prebiotic Chemistry Assemblies of L-Cysteine on Defect-Free Pyrite Terraces}

\author{
Adriana E. Candia$^{a,b,c,1}$,
Sindy J. Rodríguez$^{a,1}$,
Vanina G. Franco$^{a,d,e}$,\\
\textbf{Mario C. G. Passeggi (Jr.)}$^{a,d}$,
\textbf{Jorge Lobo-Checa}$^{f,g}$,
and \textbf{Myriam H. Aguirre}$^{f,g}$
}

\author{
Adriana E. Candia$^{a,b,c,1,*}$,
Sindy J. Rodríguez$^{a,1,*}$,
Vanina G. Franco$^{a,d,e}$,
Mario C. G. Passeggi (Jr.)$^{a,d}$,
Jorge Lobo-Checa$^{f,g}$,
and Myriam H. Aguirre$^{f,g}$
}

\affiliations{

$^{a}$ Instituto de Física del Litoral (IFIS-Litoral, CONICET--UNL), Güemes 3450, 3000 Santa Fe, Argentina

$^{b}$ Laboratorio de Microscopías Avanzadas (LMA), Universidad de Zaragoza, 50018 Zaragoza, Spain

$^{c}$ Centro de Física de Materiales (CSIC--UPV/EHU), San Sebastián 20018, Spain

$^{d}$ Departamento de Física, Facultad de Ingeniería Química, Universidad Nacional del Litoral, Santiago del Estero 2829, 3000 Santa Fe, Argentina

$^{e}$ Departamento de Física, Facultad de Bioquímica y Ciencias Biológicas, Universidad Nacional del Litoral, Ciudad Universitaria, 3000 Santa Fe, Argentina

$^{f}$ Instituto de Nanociencia y Materiales de Aragón (INMA), CSIC--Universidad de Zaragoza, 50009 Zaragoza, Spain

$^{g}$ Departamento de Física de la Materia Condensada, Universidad de Zaragoza, 50009 Zaragoza, Spain

$^{1}$ These authors contributed equally.

$^{*}$ Correspondence:
\texttt{adriana.candia@santafe-conicet.gov.ar};
\texttt{sindy.rodriguez@santafe-conicet.gov.ar}

}

\begin{document}
\maketitle
\begin{abstract}
Wächtershäuser’s  theory proposes iron-sulfur minerals as key platforms for molecular synthesis and supramolecular organization in prebiotic environments. However, defects have been traditionally considered at the center of such assemblies, thereby underestimating the contributions of regular and pristine interfaces. Here, we combine scanning tunneling microscopy and spectroscopy (STM/STS) with density functional theory (DFT) to investigate the fundamental prebiotic chemistry system of L-Cysteine (L-Cys) on defectless FeS$_2$(100) terraces. To do so, we first achieved atomically ordered, defect-free terraces that act as support of two distinct supramolecular phases of L-Cys: one compact, highly ordered supramolecular network and another less packed, labile supramolecular network. We unveil trimer-based intermolecular interactions to be at the origin of these pattern formations. These results demonstrate that L-Cys self-assemblies can be hosted on flawless FeS$_2$ terraces due to the cooperative interplay between substrate electronic structure and intermolecular interactions, without the participation of dominant defects. Therefore, the  autocatalytic activity of pyrite could have triggered the on-surface polymerization process of these non-static self-assembled structures under primordial conditions, thereby endorsing Wächtershäuser’s postulates on the origin of life.
\end{abstract}


\section{Introduction}
According to Wächtershäuser’s theory, the primordial origin of life and the emergence of prebiotic complexity was triggered by the physicochemical activity of mineral surfaces catalysing reactions from self-assembled organic building blocks. At such early stage, the abundance and stability of transition metal sulfides at the Earth’s crust could have acted as basis for the adsorption, concentration, and transformation of simple molecules under hydrothermal conditions.\cite{Wachtershauser1988, Cody2000, Russell1997, Martin2007, Seewlad2006, Door2003, Matreux2024}
Within this chemical evolution scenario, mineral surfaces must have played a fundamental role by stabilizing, concentrating, and organizing molecular species.\cite{Sutherland2017, Joyce2012, Benner2019, Nghe2025, Hazen2010, Garcia-Ruiz2020}
Under such increasingly complex chemical conditions,\cite{Smith2016,Szostak2017} iron–sulfur minerals --such as pyrite (FeS$_2$)-- could have played a key role in sight of their ability to promote organic synthesis in hydrothermal environments\cite{Cody2000, Russell1997, Martin2007}, placing these substrates as model systems capable of bridging surface science, geochemistry, and prebiotic chemistry.\cite{Vaughan1978, Rickard2007, Morse1999}

Among the possible organic precursors proposed in this chemical evolution framework, amino acids are natural candidates both because they constitute the basis of proteins and also given its expected abundance at the early Earth through both endogenous and exogenous sources.\cite{Glavin2009, Glavin2010, Zorzano2025, Mojarro2025, Sandford2025, Furukawa2026, Nguyen2025} The L-cysteine (L-Cys, SHCH$_2$CH(NH$_2$)COOH) amino acid stands out among them due to its multifunctional nature, as it combines thiol (–SH), amino (–NH$_2$), and carboxyl (–COOH) groups within a single molecule (see Fig.~\ref{fig:scheme1}), providing different protonation states depending on its chemical environment --including neutral and zwitterionic (ZW) forms--. Even in absence of explicit solvents, it is notable that surface interactions and local electrostatic fields can promote charge separation, while molecule–surface interactions may stabilize nonconventional zwitterionic states (ZW$^{\rm nc}$) through surface induced proton transfer and charge redistribution.\cite{Smith2025} This chemical adaptability enables multiple adsorption pathways, coordination to metal centers, participation in redox processes, or the formation of complex intermolecular interaction networks, \cite{Franco2024, Elina2026CysAg} thereby requiring meticulous exploration on abundant primordial mineral surfaces.

Indeed, extensive studies of organic adsorption on FeS$_2$(100) have focused on structures nucleating at defects in the form of sulfur vacancies, undercoordinated Fe atoms or step edges. However, these defect sites are known to strongly modify the local electronic structure of pyrite and introduce localized electronic states within the band gap.\cite{Herbert2013, Eggleston1996, Rosso1999b, Rosso2006} This leads to dissociative or multidentate  L-Cys configurations that are stabilized by S–Fe and O/N–Fe interactions. \cite{Smith2025, SanchezArenillas2017, DosSantos2018, GalvezMartinez2019, Rosso2006, Stirling, Nesbitt2000} The difficulty of disentangling the intrinsic role of the Fe–S interface from the effects of substrate structural heterogeneity have lead to the erroneous argument that molecular ordering on pyrite is solely based on defect-mediated processes. Such misconception can be confronted by demonstrating that organized molecular architectures can emerge directly on atomically ordered FeS$_2$(100) surfaces upon absence of dominant defects.

In this work, we contest this popular misconception by combining \textbf{STM/STS} experiments and \textbf{DFT} calculations to shed light on the adsorption and self-assembly of \textbf{L-Cys} on atomically ordered \textbf{FeS$_2$(100)} terraces. After achieving structurally and electronically well-defined defect-free surfaces, we report the supramolecular organization that directly emerges on Fe–S interfaces and identify the microscopic mechanisms underlying the observed molecular phases. Such finding go beyond primordial surface chemistry, since L-Cys is central in biological systems through the stabilization of Fe–S clusters (Fe$_2$S$_2$, Fe$_3$S$_4$, and Fe$_4$S$_4$), which are widely regarded as ancestral structural motifs in electron-transfer proteins such as ferredoxins.\cite{Beinert1997, Johnson2005, Benner2019, Hazen2010} This intimate connection between sulfur-containing amino acids and Fe–S motifs makes L-Cys a particularly relevant model system for investigating biomolecule--mineral interactions and their potential implications for prebiotic chemical evolution.

\section{Results and discussion}
\label{sec:headings}
A structurally and electronically well-defined FeS$_2$(100) surface was achieved as a defect-free reference platform for investigating intrinsic molecule--surface interactions (see Materials and Methods). Its structural and electronic properties were characterized by scanning tunneling microscopy and spectroscopy (STM/STS) under ultra-high vacuum (UHV) and cryogenic conditions (4.8 K). On this substrate, the adsorption of L-Cys on defect-free terraces was investigated by combining STM/STS measurements with density functional theory (DFT) calculations, enabling direct correlation between the atomic structure, stability, and electronic properties of the observed phases.

\begin{figure}
\centering
\includegraphics[width=0.5\linewidth]{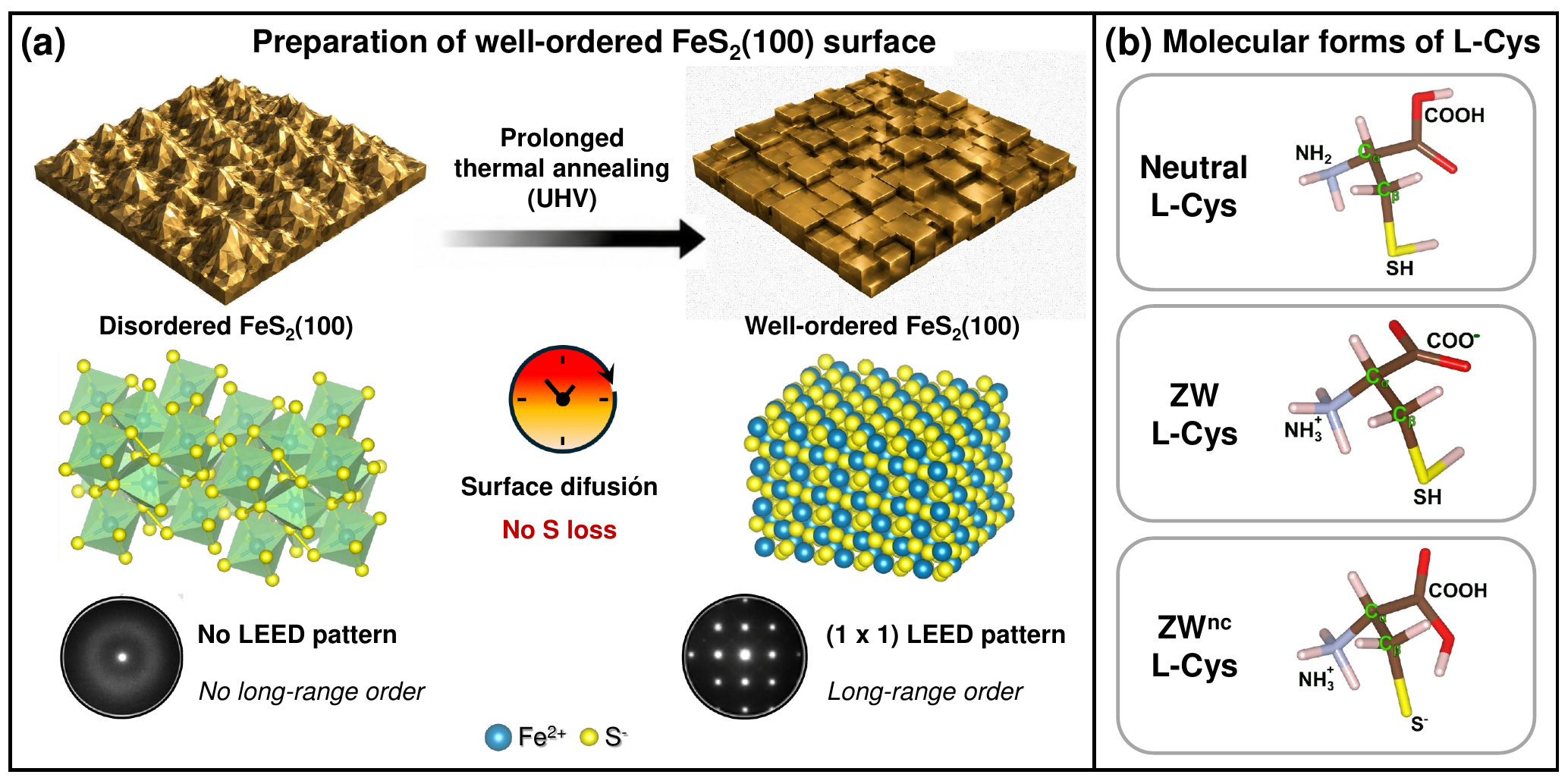}
\caption{(a) Schematic illustration of a well-ordered FeS$_2$(100)-(1$\times$1) surface under UHV conditions. Prolonged annealing promotes surface ordering while preserving sulfur stoichiometry, yielding a bulk-terminated surface with long-range order. (b) Molecular configurations of L-cysteine (L-Cys) considered in this work: neutral, zwitterionic (ZW), and non-conventional zwitterionic (ZW$^{\rm nc}$).}
\label{fig:scheme1}
\end{figure}

Figure~\ref{fig:scheme1} summarizes the surface preparation procedure and the molecular configurations of L-Cys considered in this study. As illustrated in (a), repeated Ar$^+$ sputtering followed by prolonged annealing under UHV promotes surface ordering while preserving the intrinsic symmetry of FeS$_2$(100), as confirmed by LEED. Panel (b) shows the three L-Cys configurations analyzed: the neutral form, the conventional zwitterionic (ZW) form, where a proton is transferred from the --COOH to the --NH$_2$ group, and a non-conventional zwitterionic configuration (ZW$^{\rm nc}$), involving proton transfer from the --SH to the --NH$_2$ group.

\subsection{Pristine FeS$_2$(100): structural and electronic reference conditions}
\begin{figure}[h!]
\centering
\includegraphics[width=0.5\linewidth]{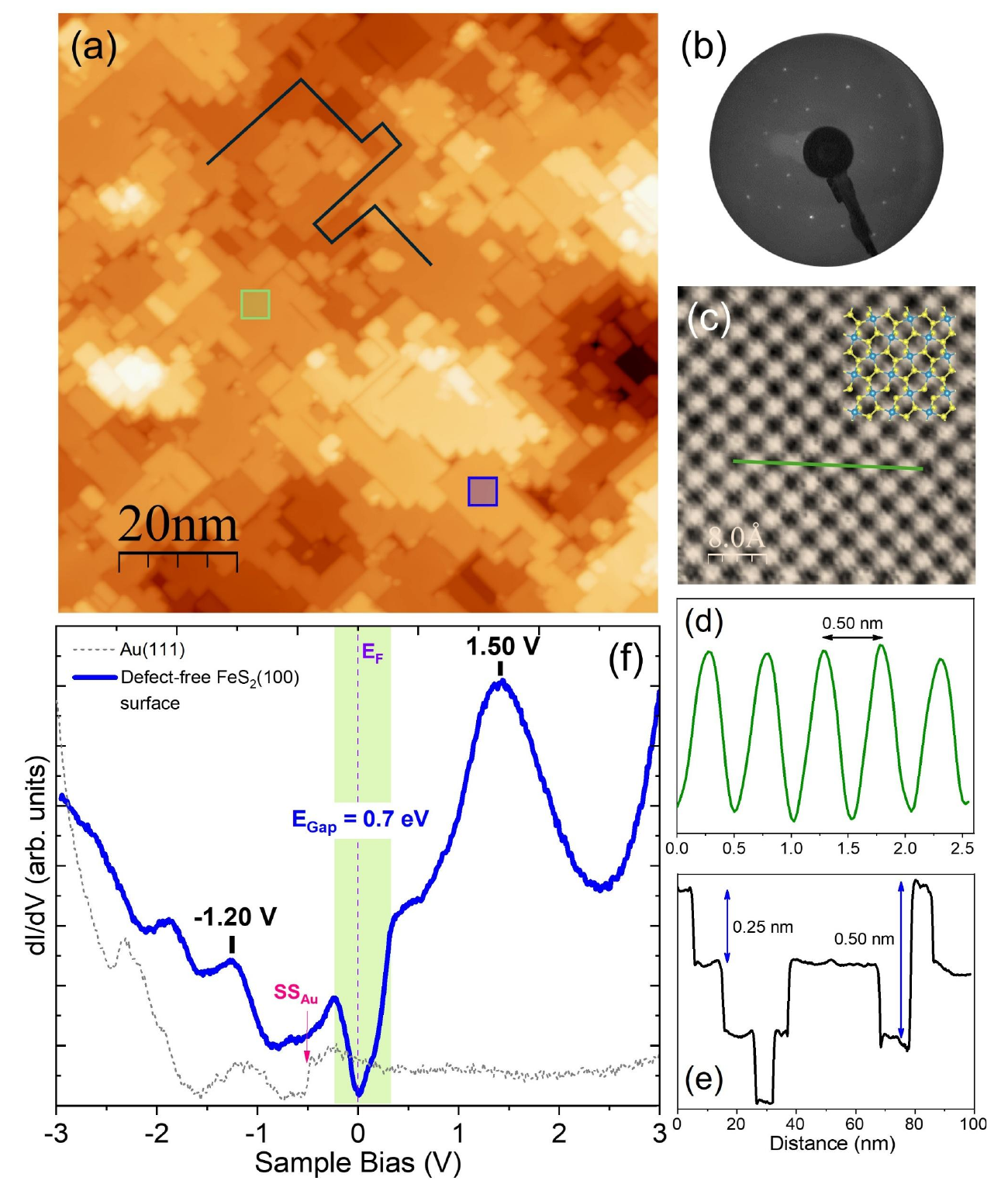}
\caption{\textbf{Structural and electronic characterization of the defect-free FeS$_2$(100) surface.} (a) Large-scale STM topographic image acquired after annealing at 590~K for 6 h under UHV conditions. (b) (1 $\times$ 1) LEED pattern recorded at 110 eV. (c) Atomically resolved STM image acquired within the green square marked in (a), with the structural model overlaid as a visual guide (Fe: blue; S: yellow). (d) Line profile measured along the green line in (c), showing the 0.50 nm surface periodicity. (e) Height profile measured along the blue path in (a), revealing step heights of 0.25 and 0.50 nm. (f) Local dI/dV spectrum acquired within the blue square in (a). The FeS$_2$(100) spectrum (blue) exhibits a surface band gap of $\sim$0.70 eV, bounded by states at approximately $-0.25$ V and $+0.45$ V, together with occupied and unoccupied features centered at $-1.20$ V and $+1.50$ V, respectively. The dotted gray curve corresponds to Au(111), used as a tip reference prior to the pyrite measurements, exhibiting the characteristic Shockley surface state at $\approx -0.48$ eV. STM parameters: (a) $V_s/I_s=-2.0$~V$/50$~pA and (c) $V_s/I_s=-1.5$~V$/200$~pA.  STS parameters: Bias voltage modulation of 10 mV$_{rms}$ at 817.3 Hz with the following setpoints: $V_s/I_s=-1.0$~V$/50$~pA.}
\label{fgr:STM_Clean_Py}
\end{figure}

The surface defects of FeS$_2$(100) --S vacancies or undercoordinated Fe sites-- introduce localized electronic states within the band gap that modify the molecule–surface interaction and act as preferential adsorption and nucleation centers.\cite{Herbert2013, DeLeeuw2000, Arrouvel2018} Thus, it is essential to start from a structurally and electronically well-defined substrate to access the intrinsic properties of the Fe–S interface (see experimental section in the SI).
Figure~\ref{fgr:STM_Clean_Py} correlates the UHV prepared surface morphology with the local electronic response of FeS$_2$(100). The large-scale STM image in Figure~\ref{fgr:STM_Clean_Py}(a) reveals atomically flat terraces separated by straight step edges aligned along the main crystallographic directions, evidencing the high structural quality of the substrate. The LEED (1$\times$1) pattern in Figure~\ref{fgr:STM_Clean_Py}(b) confirms the surface quality and long-range order. At the atomic scale, the lateral periodicity of 0.50 nm observed in Figure~\ref{fgr:STM_Clean_Py}(c) matches the Fe–Fe distance predicted by the structural model, while the height profile in Figure~\ref{fgr:STM_Clean_Py}(e) reveals step heights of 0.25 and 0.50 nm, corresponding to half and one pyrite unit cell, respectively. Comparison with simulated STM images of ideal and defective surfaces \textbf{(Figure~S1 in the SI)} confirms the defect-free nature of the prepared FeS$_2$(100) surface.

The local electronic structure, probed by dI/dV spectroscopy and shown in Figure~\ref{fgr:STM_Clean_Py}(f), exhibits a pronounced suppression of the density of states near the Fermi level, between approximately $-0.25$ V and $+0.45$ V, resulting in an effective surface band gap of $\sim$0.70 eV, smaller than the reported bulk value of $\sim$0.95 eV.\cite{Herbert2013,Rosso1999a} This reduction is attributed to the modified coordination environment of surface Fe atoms, which alters the local crystal field and gives rise to Fe 3d-derived surface states. Two prominent features are observed at approximately $-1.20$ V and $+1.50$ V, consistent with previous STS studies of pyrite.
\cite{Herbert2013,Rosso1999b,Schaufuss1998} The reproducibility of these spectral features across different terraces, together with the absence of localized in-gap states in the STS line scans shown in \textbf{Figure S2}, confirm the electronic homogeneity of the substrate. The conductance suppression around zero bias may be partially influenced by tip-induced band bending (TIBB), which locally modifies the electrostatic potential and reduces the tunneling probability near the Fermi level.\cite{Feenstra1994,Teichmann2008} Consistently, DFT-derived LDOS calculations predict that surface defects would introduce in-gap states and modify the spectral response, which are not visible in our experimental observations (\textbf{Figures~S3 and S4}). Thus, the prepared FeS$_2$(100) surface is structurally ordered and electronically homogeneous, providing a reliable defect-less reference for molecular assemblies.

\begin{figure*}[t!]
\centering
\includegraphics[scale=0.65]{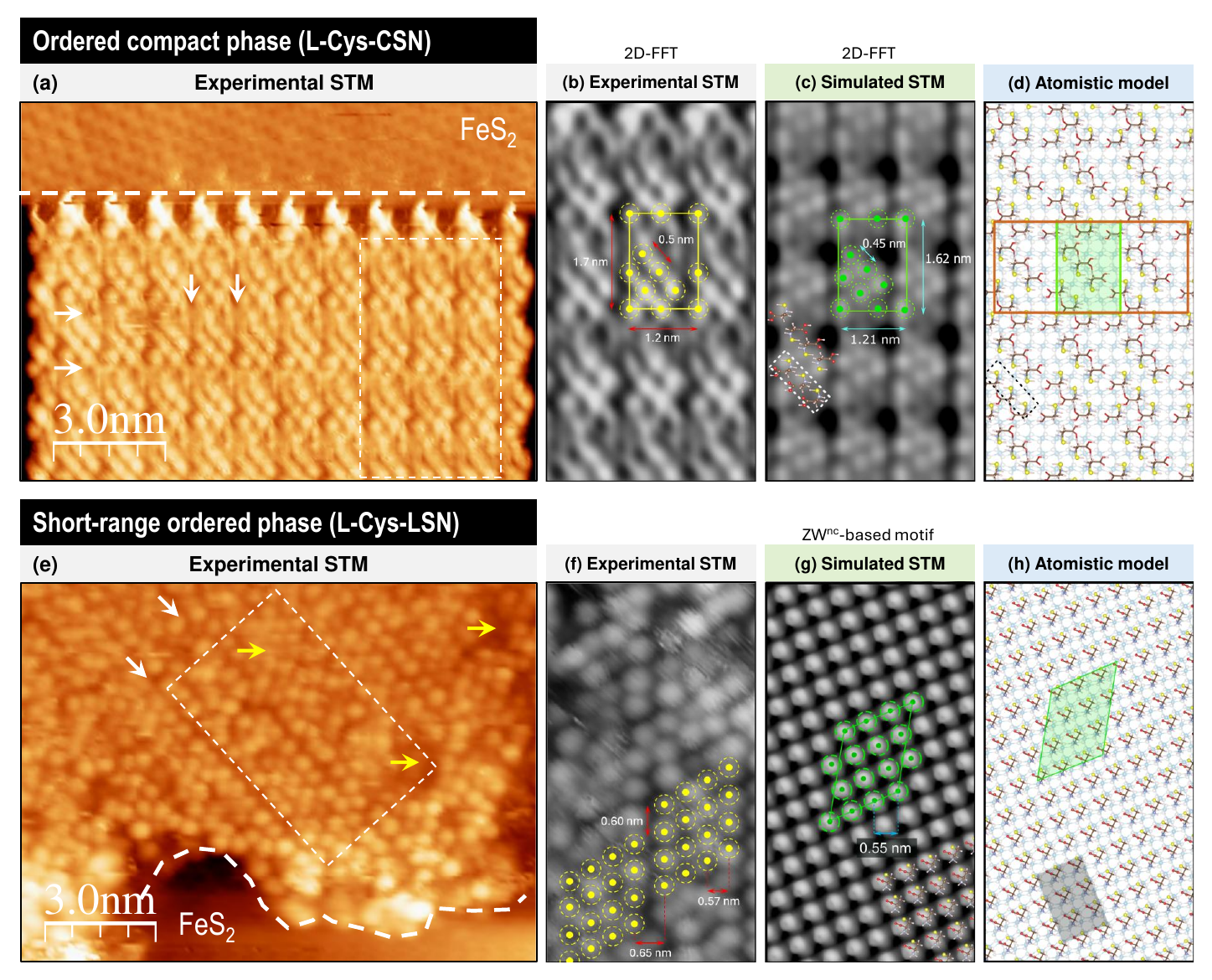}
\caption{\textbf{Experimental and theoretical characterization of the L-Cys supramolecular phases formed on FeS$_2$(100).} The upper panels correspond to the ordered compact supramolecular network (L-Cys-CSN). (a) STM topographic image of a representative L-Cys-CSN domain. (b) High-resolution STM image together with the corresponding two-dimensional fast Fourier transform (2D-FFT) analysis of the region delimited by the dashed box in (a), revealing an approximately rectangular periodicity with lattice parameters of $a \approx 1.2$ nm and $b \approx 1.7$ nm. Dashed yellow circles are included as a visual guide to the experimentally resolved molecular arrangement. (c) Simulated STM image obtained from the DFT-optimized model and its corresponding 2D-FFT analysis, reproducing the experimental periodicities. (d) Atomistic model of the L-Cys-CSN phase on FeS$_2$(100). The green region indicates the proposed unit cell, while the orange rectangle delineates the $7 \times 3$ commensurate surface supercell employed in the simulations. The lower panels correspond to the labile supramolecular network (L-Cys-LSN). (e) STM topographic image of a second molecular phase characterized by a less compact organization and short-range order. White arrows indicate preferred correlation directions, whereas yellow arrows highlight local variations in molecular packing. (f) High-resolution STM image of the region delimited by the dashed box in (e), showing locally resolved molecular entities and characteristic intermolecular separations of 0.55–0.65 nm. Dashed yellow circles are included as a visual guide. (g) Simulated STM image based on the supramolecular motif derived from the ZW$^{\rm nc}$ configuration, reproducing the intermolecular spacing and local organization observed experimentally. Dashed green circles highlight the proposed molecular arrangement. (h) Atomistic model associated with the L-Cys-LSN phase, where the green region delineates the representative local packing configuration employed in the simulations. Experimental STM parameters: (a)–(b) $V_s/I_s=-1.20$~V$/50$~pA; (e)–(f) $V_s/I_s=-2.0$~V$/50$~pA. Simulated STM images in panels (c) and (g) were generated within the Tersoff–Hamann approximation using an isovalue of $1\times10^{-5}$ and bias voltages of $-1.20$ V and $-2.0$ V, respectively.}
\label{fgr:STM_islands}
\end{figure*}

\subsection{Self-assemblies of L-Cys on a FeS$_2$(100) defect-free surface}

Figure~\ref{fgr:STM_islands} shows the two supramolecular phases identified upon adsorption at RT of L-Cys on FeS$_2$(100). Both networks form these aggregates on the substrate terraces without preferential nucleation at step edges or surface defects (\textbf{Figure S5 in the SI}), suggesting that their self-assembly is governed by the cooperative interplay between the structural and electronic templating imposed by the interface and the intermolecular interactions.\cite{Suzuki2019} These observations demonstrate that a single molecule--substrate interface can stabilize more than one supramolecular arrangement.

Figure~\ref{fgr:STM_islands}(a) shows the most compact and structurally robust phase, denoted as the L-Cys compact supramolecular network (L-Cys-CSN). These domains exhibit straight edges, highlighted by the white dashed outline, a well-defined internal geometry, and high stability under repeated STM scanning. The white arrows indicate the preferred correlation directions, revealing a highly ordered periodic arrangement. The experimental FFT shown in Figure~\ref{fgr:STM_islands}(b) defines an approximately rectangular unit cell of $\sim$1.2 $\times$ 1.7 nm$^2$, while the superimposed yellow circles serve as a visual guide to the periodically resolved features and their correspondence with the simulated STM images and atomistic model shown in Figure~\ref{fgr:STM_islands}(c,d). The excellent agreement between experiment and simulation indicates that the proposed structure accurately describes the supramolecular organization of this phase.

In contrast, Figure~\ref{fgr:STM_islands}(e) shows a second morphology, denoted as the L-Cys labile supramolecular network (L-Cys-LSN), characterized by larger, more irregular domains and lower stability when scanned under the STM tip. The yellow arrows highlight regions where bright features are displaced during image acquisition, evidencing enhanced local mobility, whereas the white arrows indicate short-range spatial correlations within the network. Comparison between the experimental STM images and the simulated structures (Figure~\ref{fgr:STM_islands}(f--h)) reveals that, although the agreement is less direct than for the compact phase, the proposed model reproduces the essential structural features observed experimentally. This behavior is consistent with the more dynamic nature of L-Cys-LSN, where local fluctuations hinder the development of a unique long-range periodic arrangement. A detailed atomistic interpretation of this phase is provided in the following section based on DFT calculations.

A quantitative analysis of the intradomain organization reveals clear differences between the two supramolecular networks. While L-Cys-CSN exhibits a homogeneous topography and a lower apparent height ($\sim$0.20 nm), consistent with a compact and ordered packing arrangement, L-Cys-LSN displays a larger apparent height ($\sim$0.30 nm) and a more heterogeneous topographic contrast, indicative of a more open organization and greater conformational freedom of the adsorbed molecules. Representative line profiles and the corresponding structural parameters are presented in \textbf{Figure S5 of the SI}.

\subsection{Adsorption geometries and functional group orientation of L-Cys on FeS$_2$(100)}

\begin{figure*}[h!]
 \centering
 \includegraphics[height=10.0cm]{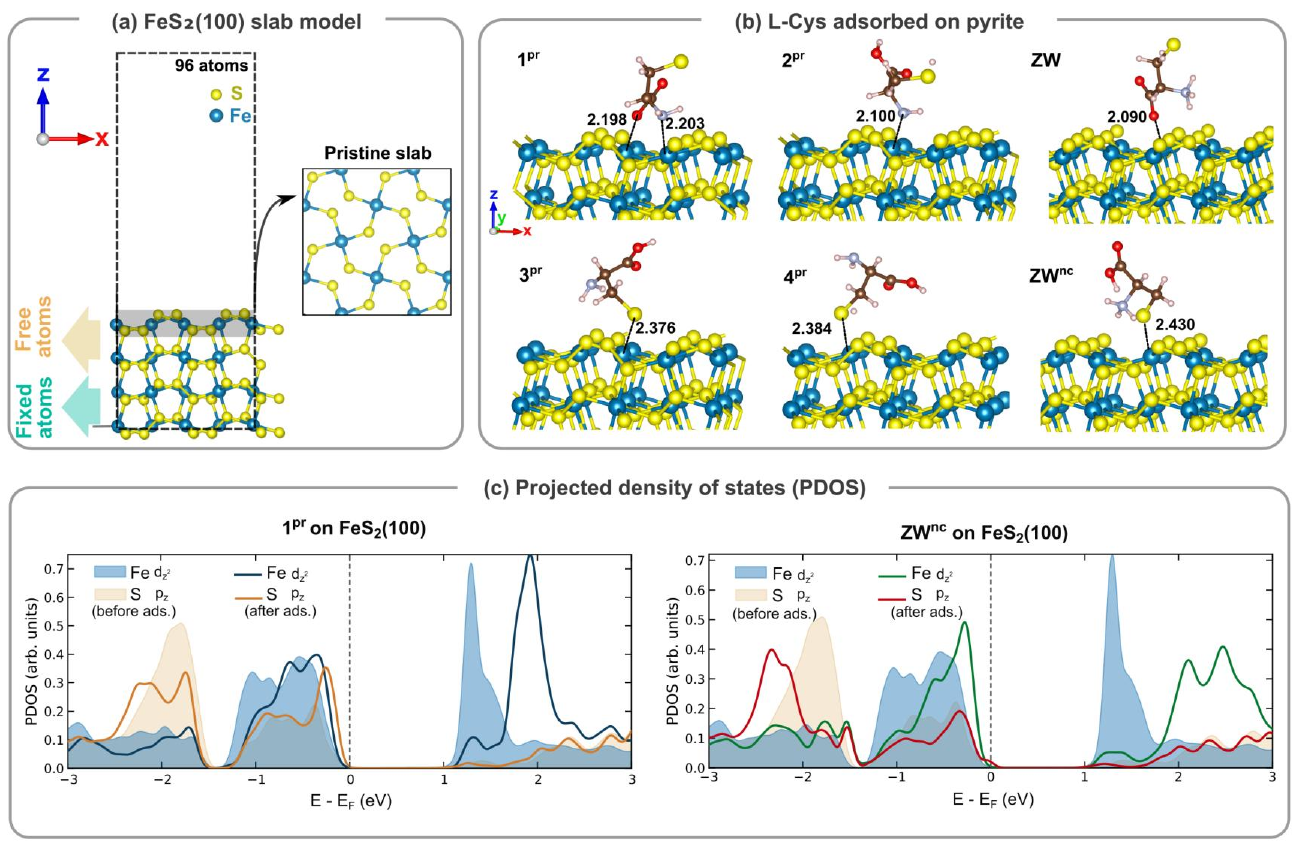}
 \caption{\textbf{DFT models of L-Cys adsorption on the pristine FeS$_2$(100) surface.} 
(a) Supercell used to model the sulfur-terminated FeS$_2$(100) surface, showing the slab geometry and the division between relaxed and fixed atomic layers. (b) Optimized adsorption configurations considered in this work: four neutral geometries (1$^{\rm pr}$–4$^{\rm pr}$), a conventional zwitterionic structure (ZW), and a non-conventional zwitterionic configuration (ZW$^{\rm nc}$) (see Table S1). Relevant molecule--surface bond distances are indicated. The most stable structure is ZW$^{\rm nc}$, in which the deprotonated sulfur atom binds to a surface Fe atom while the carboxyl group points away from the surface. (c) Projected density of states (PDOS) of representative adsorption configurations compared with the clean surface, highlighting the Fe 3d$_{z^{2}}$ and S 3p$_z$ contributions before and after adsorption.}
 \label{fgr:super_cell}
\end{figure*}

The previous atomistic analysis of the two coexisting supramolecular phases that exhibit distinct degrees of compactness, internal order, and mechanical stability, can be used to establish the adsorption geometries and intermolecular interactions responsible for the stabilization of both networks. To determine the preferred adsorption sites of L-Cys on the experimentally observed defect-free FeS$_2$(100) surface, DFT calculations were performed exploring different molecular orientations and adsorption distances relative to the surface. All atomic positions of the molecule and the upper pyrite layers were fully relaxed during the geometric optimization process, see Figure~\ref{fgr:super_cell}(a). In addition to neutral adsorption geometries, zwitterionic (ZW) configurations were also considered, since L-Cys may adopt this form under appropriate conditions.\cite{Lehninger2017} Although the present model does not include an explicit solvent, the surface may stabilize separated charges. Accordingly, two ZW configurations were explored: a conventional structure, in which the carboxyl group is deprotonated while the amino group remains protonated, and an unconventional zwitterionic configuration (ZW$^{\rm nc}$),\cite{Smith2025} arising from intramolecular proton transfer from the thiol group to the amino group.

Figure~\ref{fgr:super_cell}(b) shows side views of the optimized adsorption structures obtained after structural relaxation, while the corresponding adsorption energies and molecule--surface distances are summarized in  \textbf{Table S1}. Among all the explored geometries, the most stable configuration corresponds to ZW$^{\rm nc}$, followed by the neutral (1$^{\rm pr}$) configuration. In the 1$^{\rm pr}$ adsorption geometry, the oxygen (O) atom of the carboxyl group and the nitrogen (N) atom of the amino group simultaneously coordinate with surface Fe atoms, forming a bidentate interaction with the substrate. The optimized Fe--O and Fe--N bond lengths are 2.198~\AA\ and 2.203~\AA, respectively, consistent with values reported for amino acid adsorption on iron-containing surfaces. \cite{Zheng2018} In this configuration, both heteroatoms are oriented toward the surface, while the sulfur (S) atom of the thiol group remains exposed toward the vacuum.

In contrast, the most stable configuration, corresponding to ZW$^{\rm nc}$, exhibits a markedly different adsorption geometry. The intramolecular hydrogen (H) transfer from the thiol group to the amino group generates a deprotonated sulfur center that directly coordinates with a surface Fe atom, forming a stable Fe--S bond that anchors the molecule to the substrate. As a consequence, the carboxyl group becomes oriented toward the vacuum, highlighting the ability of the surface to stabilize proton-redistributed molecular states even in the absence of solvent.

We computed the projected density of states (PDOS) of the surface atoms involved in adsorption for the two most stable molecular configurations (1$^{\rm pr}$ and ZW$^{\rm nc}$; see Figure~\ref{fgr:super_cell}c). For the neutral configuration, adsorption significantly modifies both the Fe 3d$_{z^2}$ and S 3p$_z$ states near the Fermi level, particularly in the unoccupied region, indicating a relatively delocalized molecule--surface interaction. In contrast, the electronic perturbation in the ZW$^{\rm nc}$ configuration is dominated by the Fe 3d$_{z^2}$ states, while the S 3p$_z$ states remain largely unaffected, suggesting a more localized Fe-centered interaction. These differences are consistent with the experimentally observed supramolecular phases, where the delocalized interaction of the neutral configuration favors cooperative molecular organization, whereas the localized interaction associated with ZW$^{\rm nc}$ correlates with weaker lateral coupling and the absence of long-range order.

\subsection{Atomistic models of the self-assembled structures of L-Cys on FeS$_2$(100)}
To further correlate these geometries with the experimental STM observations, simulated STM images within the Tersoff–Hamann approximation at the experimental voltages were computed for isolated L-Cys adsorbates using the 1$^{\rm pr}$ and ZW$^{\rm nc}$ configurations (\textbf{Figure S7 in the SI}). Although the simulations reproduce the local appearance of individual molecular features, neither configuration accounts for the rectangular periodicity observed in the ordered L-Cys-CSN domains. This indicates that such structures do not originate from isolated adsorbates, but rather from intermolecular organization within the molecular overlayer. Indeed, for large molecular aggregates, intermolecular interactions---particularly H-bonding between neighboring molecules---become dominant. Although the ZW$^{\rm nc}$ is energetically more stable for isolated adsorption, these interactions can alter the subtle balance and stabilize assemblies based on the neutral geometry. Consequently, both configurations (1$^{\rm pr}$ and ZW$^{\rm nc}$) must be considered as non-isolated structural building blocks for constructing the aggregates experimentally observed.

To establish the structural origin of the observed L-Cys-CSN phase we must first construct atomistic models containing several units based on the previously identified adsorption geometries. The STM images reveal ordered domains with an approximately rectangular periodicity of $a \approx 1.2$ nm and $b \approx 1.7$ nm (Figure~\ref{fgr:STM_islands}(a,b)), together with compact arrays of bright features of similar apparent height, highlighting the periodicity of the supramolecular motifs. Similarly to isolated molecules, dimer-like assemblies based on the neutral 1$^{\rm pr}$ and ZW$^{\rm nc}$ configurations fail to reproduce the experimental STM image contrast (\textbf{Figures S7--S10 in the SI}). Therefore, we need to resort to trimer-based configurations as the minimal structural motif to accurately reproduce the experimental intermolecular spacings and topographic contrast (Figure~\ref{fgr:STM_islands}(c,d)). 

These simulated structures of the L-Cys-CSN phase, requires that three L-Cys molecules occupy symmetry-related adsorption sites with the thiol groups oriented toward the vacuum. A commensurate 7 $\times$ 3 FeS$_2$(100) surface supercell (orange box in Figure \ref{fgr:STM_islands}(d)) reproduces the extended experimentally observed periodicity. The unit cell consists of pairs of trimers arranged in a face-to-face configuration together with an additional linker molecule located in the central region (green box in Figure \ref{fgr:STM_islands}(c)) (see \textbf{Section 7 of the SI} for additional structural details). The simulated images results in intermolecular spacing of $\sim$0.45–0.50 nm and periodicities of $a \approx 1.21$ nm and $b \approx 1.62$ nm, which are in close agreement with the experiment (see \textbf{Figure S11} for representative simulated STM images and  corresponding 2D-FFT).
 Within this Tersoff–Hamann approximation, the STM contrast is dominated by sulfur-derived occupied states associated with the thiol groups, explaining the bright features observed in both experimental and simulated images. In essence, the L-Cys-CSN domains is unveiled as a supramolecular arrangement of neutral adsorption geometries organized into trimer-based motifs, where molecule–surface interactions and intermolecular hydrogen bonding collectively define the periodic architecture observed on the FeS$_2$(100) surface.

In contrast, the lower packing density structure of L-Cys-LSN increases the intermolecular separations to $\sim$0.55–0.65 nm (Figures \ref{fgr:STM_islands}(e,f)), which reduces the intermolecular interactions and prevents the formation of extended periodic arrangements compared to the L-Cys-CSN phase. STM simulations performed for trimeric assemblies provide such weakly packed arrangements in the ZW$^{\rm nc}$ adsorption configuration (Figure \ref{fgr:STM_islands}(g)), where Fe-S interactions dominate the surface anchoring and preferentially expose the COOH groups toward the vacuum. Consequently, the intermolecular coupling is reduced precluding compact supramolecular motifs. 
In this way, the calculated STM images reproduce the experimentally observed local arrangement and characteristic intermolecular spacing along the principal packing direction with differences in the orthogonal direction. Such differences are consistent with the absence of well-defined two-dimensional periodicity in the experimental L-Cys-LSN domains, which suggests that these disordered regions originate from locally stable ZW$^{\rm nc}$-based assemblies that remain weakly packed and fail to develop extended supramolecular order.

\section{Conclusions}
In this work, we combined STM, STS, and DFT calculations to investigate the adsorption and supramolecular organization of the fundamental prebiotic chemistry system of L-Cysteine (L-Cys) on FeS$_2$(100). We achieved a defect-free FeS$_2$(100) surface that exhibits homogeneous structural and electronic character, which provides an ideal platform for primordial chemistry studies. Upon L-Cys deposition, we report the spontaneous formation of two distinct supramolecular phases on the FeS$_2$(100) terraces through the cooperative interplay between molecule–surface interactions and intermolecular hydrogen bonding. DFT calculations combined with STM simulations show that isolated adsorption geometries and simple dimer-like assemblies cannot reproduce the experimental complexity. Instead, trimer-based motifs emerge as the minimal structural units to unveil the compact L-Cys-CSN phase, whereas weaker adsorption configurations give rise to loosely packed assemblies consistent with the L-Cys-LSN phase. The coexistence of these structures originates from distinct adsorption regimes associated with different protonation states of L-Cys and their corresponding electronic coupling to the FeS$_2$(100) surface.

More broadly, our results demonstrate that supramolecular organization on pyrite does not require defect-mediated templating, but can emerge directly from structural and electronically homogeneous pristine Fe–S surfaces. 
Therefore, defect-free FeS$_2$(100) can act as an active playground for directing collective supramolecular networks with tailored structural and electronic properties and foreseen impact in  hybrid organic–inorganic interface design.
Finally, in the context of prebiotic chemistry, these results provide experimental evidence that well-ordered iron–sulfur surfaces could have promoted molecular organization under early-Earth conditions, supporting mineral-assisted scenarios for chemical evolution.

\subsubsection*{Experimental details}
Samples were prepared from natural FeS$_2$ crystals (Minas de Navajún, Spain), exposing the (100) surface. After chemical cleaning, the samples were transferred into a UHV chamber ($\sim 1 \times 10^{-10}$ mbar). Ordered FeS$_2$(100) surfaces were obtained by repeated Ar$^+$ sputtering cycles (500 eV) followed by annealing at 590~K for 6 h, promoting surface reorganization without significant S loss. Long-range order was verified by low-energy electron diffraction (LEED). Commercial L-Cys ($\geq 98\%$ purity) was deposited under UHV using a home-built evaporator equipped with a resistively heated quartz crucible and a quartz crystal microbalance. After degassing, L-Cys was sublimated at 390~K onto FeS$_2$(100). Low-temperature STM/STS measurements were performed at 4.8 K using a Scienta Omicron LT-qPlus microscope and W tips. Differential conductance ($dI/dV$) spectra were acquired using lock-in detection. STM images were recorded in constant-current mode and processed with WS$\times$M software. 
\subsubsection*{Computational Details}
First-principles calculations were performed within density functional theory (DFT) using the OpenMX 3.9 code within the GGA-PBE approximation, including spin polarization, DFT+$U$, and dispersion corrections through the DFT-D3 scheme. The FeS$_2$(100) surface was modeled using sulfur-terminated slabs corresponding to the most stable structures reported in the literature. Structural optimizations were carried out using periodic supercells separated by a vacuum region. Additional experimental procedures, computational parameters, structural models, and extended theoretical analyses are provided in the Supporting Information.

\subsubsection*{Acknowledgment}
The authors acknowledge Carlos Martín-Sacristán for continuous technical support, as well as the use of the Servicio General de Apoyo a la Investigación (SAI) and the Laboratorio de Microscopías Avanzadas (LMA, Universidad de Zaragoza). This work was supported by the Spanish Ministry of Science and Innovation (PID2022-138750NB-C21, MCIN/AEI/10.13039/501100011033 and ERDF “A way of making Europe”), the Government of Aragón (E12-23R), the Severo Ochoa Programme for Centres of Excellence in R\&D (CEX2023-001286-S), the European Research Council through the Marie Skłodowska-Curie Action ULTIMATE-I (No. 101007825) and NESTOR (No. 101007629), the IKUR Strategy through the collaboration agreement between the Ikerbasque Foundation and the Materials Physics Center on behalf of the Basque Government, CONICET–ANPCyT (PICT2021-INVI-00863, PICT2019-04545), CONICET (PIP-2021-101517), UNL (CAI+D50620190100016LI and CAI+D85520240100061LI), and ASaCTei (PEICID-2022-072 and PEICA-2023-039). Computational resources were provided by the Pirayú cluster (ASaCTei, Argentina; AC-00010-18), part of Argentina’s National High Performance Computing System, and by CESAR at the BIFI Institute, UNIZAR.

\bibliographystyle{unsrt}  
\bibliography{references}

\end{document}